\documentclass[useAMS,usenatbib]{mn2e}

\def\chan{{\it Chandra}}
\def\xmm{{\it XMM-Newton}}

\def\xh{{X$_H$}\ }

\def\efe25{{$(e~+~Fe~XXV) \rightarrow Fe~XXIV$}\ }
\def\fe25{{Fe~XXV}}
\def\ka{{K$\alpha$}\ }

\def\xmm{{\it XMM-Newton}\ }

\def\eion{{(e~+~ion)}\ }
\def\te{{T$_e$}}
\def\ne{{N$_e$}}
\def\en{{$n$}\ }
\usepackage[dvips]{graphics}
\usepackage[dvipsnames,usenames]{color}

\title{The 6.7 keV K$\alpha$ complex of He-like iron in transient
plasmas}
\author[Justin Oelgoetz and Anil K. Pradhan]
       {Justin Oelgoetz$^{1,2}$ and Anil K. Pradhan$^3$\\
       $^1$ GRA Program, Los Alamos National Laboratory (X-5), Los Alamos, NM, 87545, USA \\
       $^2$ Department of Chemistry, $^3$ Department of Astronomy,
 The Ohio State University, Columbus, OH 43210, USA}
\date{Accepted  xxxxxx 
      Received xxxxxx;
      in original form xxxxxx}

\pagerange{\pageref{firstpage}--\pageref{lastpage}}
\pubyear{2004}

\def\LaTeX{L\kern-.36em\raise.3ex\hbox{a}\kern-.15em
    T\kern-.1667em\lower.7ex\hbox{E}\kern-.125emX}

\begin{document}

\setcounter{page}{0}
\thispagestyle{empty}
\includegraphics{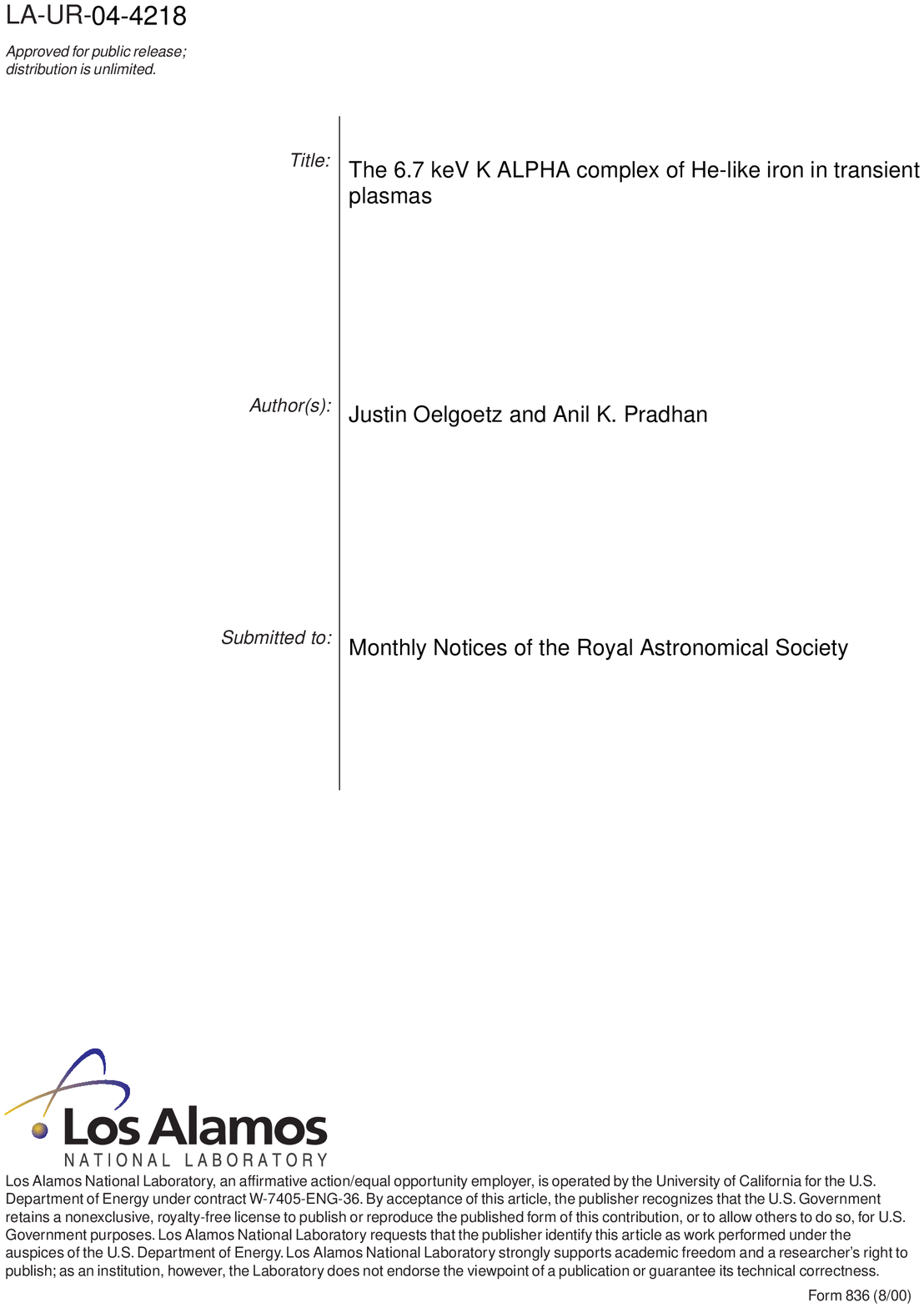}

\maketitle

\label{firstpage}

\begin{abstract}
Time dependent numerical simulations of the \ka complex of \fe25~are carried 
out as a function of temperature/density/radiation field variations in high-temperature 
astrophysical and laboratory plasmas. In addition to several well known 
features, the transient and steady state spectra reveal the effects due to (a) 
time-dependent thermal and non-thermal radiation fields, (b) photo and 
collisional excitation and ionization, and (c) high densities, on the 
`quartet' of principal w,x,y,z lines, and dielectronic satellites. The highly
detailed models show precisely how,  
assuming a temporal-temperature correlation, the X-ray intensity varies between 6.6 - 6.7 keV and undergoes a `spectral inversion' in the w and z line intensities, characterizing an ionization or a recombination dominated plasma. The dielectronic satellite intensities are the most temperature dependent features, but insensitive to density variations, and significantly contribute to the \ka complex for T $<$ 6.7 keV leading to asymmetric profiles. The 6.7 keV \ka complex should be a potential diagnostic of X-ray flares in AGN, afterglows in gamma-ray bursts, and other non-equilibrium sources with the high-resolution measurements possible from the upcoming mission {\it Astro-E2}. It is also shown that high electron densities attenuate the line intensities in simulations relevant to laboratory plasmas, such as in inertial confinement fusion, laser, or magnetic Z-pinch 
devices. 

\end{abstract}


\section{INTRODUCTION}
 
The Fe \ka spectral complex is the most prominent spectral feature in a variety of sources observed with X-ray observatories such as ASCA, \chan, and \xmm. Whereas the singular 6.4 keV feature is thought to arise from a relatively `cold' plasma through K-shell ionization by hard X-rays and subsequent fluorescence in closed $2p$-subshell systems (Fe~I~--~XVI), the higher energy features at 6.7 and 6.9 keV would arise from `hot', highly ionized He-like Fe~XXV and H-like Fe~XXVI.  Other ionization species of Fe may also contribute in between these energy ranges. It is well known that in the highly ionized regime the He-like spectrum often dominates X-ray spectra since the closed K-shell survives in the plasma over the largest temperature range and as such is a powerful diagnostic of temperature, density, and ionization state \citep{1969Gabriel,1978Mewe-a,1978Mewe-b,1981Pradhan-a,1982Pradhan,1985Pradhan-a,2000Liedahl, 2000Porquet,2000Bautista,2002Ness}. The He-like spectra are of equal interest in laboratory plasmas such as tokamaks, magnetic Z-pinch and inertial fusion devices where non-equilibrium, time-dependent, and non-thermal conditions prevail \citep[e.g.][]{1985Lee,1986Lee,2002Bailey}. 

In astrophysical situations the variation of the continuum and the line flux reveals information on the intrinsic nature of the source. For example,  \citet{2003Yaqoob} have reported time-resolved \chan  \ and {\it Rossi Timing X-ray Explorer} (RXTE) detection of Fe~XXV and Fe~XXVI \ka complexes from the narrow-line Seyfert 1 galaxy NGC 7314 and found that the ionization states of Fe respond to continuum variations.  While the ambient plasma may still be
in quasi-equilibrium, the flare model for the irradiated accretion disc is invoked to explain the Fe \ka feature and related variations in X-ray and UV fluxes \citep{2003Pounds,2003Collin}. Another example is that of time-dependent X-ray afterglows in gamma-ray bursts showing Fe \ka features \citep[e.g][]{2000Rees}.  A flare or a burst might ionize through a number of ionization states, and since recombination time-scales are much longer than the duration of the event itself the spectrum should reveal the state of ionization.  The models must therefore consider all ionization states, particularly those likely to dominate under most conditions. 

 Several recent observations of variable spectra of the iron \ka complex in AGN show that it is primarily due to \fe25~\citep{2004Iwasawa}.  The \xmm observations of the prototypical narrow-line Seyfert 1 galaxy I Zw 1 show an X-ray flare and spectral variability with both the Fe XXV \ka feature and the K edge \citep{2003Dewangan}. While currently observed \fe25~spectra are unresolved, and unable to constrain the associated parameters precisely, a more detailed analysis based on higher resolution measurements than possible from \chan ~or \xmm might be possible in the near future. Such a possibility should be afforded by the upcoming mission {\it Astro-E2} which aims to resolve the \ka complex of \fe25. The present work aims to facilitate such studies.  
 
More generally, astrophysical plasmas with He-like iron vary widely.  In the solar corona one has the characteristic `coronal' or collisional equilibrium where photoionization may be neglected. The standard coronal spectrum of Fe~XXV in equilibrium at different temperatures was reported in an earlier work, including the principal and dielectronic satellite lines and relative contributions to the \ka complex \citep{2001Oelgoetz}.  The present work is a more extensive study of the time/density dependence of a high temperature plasma as reflected in spectral variations {\it within} the 6.7 keV complex of Fe~XXV. While all ionization states of Fe are included in the time-dependent simulations, we concentrate on the detailed features of Fe~XXV since it is likely to be the prime contributor to Fe \ka in any highly ionized plasma. The results are obtained using a new code HELINE designed to study He-like ions in general employing state-of-the-art atomic data \citep{2001Oelgoetz}. A comparison of the non-equilibrium spectra is made with both collisional and/or photoionization equilibrium assuming different electron densities and radiation fluxes. Ionization fractions and corresponding spectra are computed with time-temperature profiles related to observed sources.

\section{THEORY}

 The theory of spectral formation by He-like ions has been developed by \citet{1969Gabriel,1972Blumenthal,1977Lin-b,1978Mewe-a,1981Pradhan-a,1982Bely-Dubau}. These studies assumed stationary ionization equilibrium conditions. Emission from non-stationary transient plasmas was considered by \citet{1978Mewe-b,1980Mewe}; however, atomic data employed in their work has been superseded by more recent calculations (described later). In particular, the excitation rate coefficients used in their work were computed without the important effect of resonances on cross sections. Employing improved excitation rate coefficients for He-like ions including resonances in the close coupling formulation \citep{1981Pradhan-b,1981Pradhan-c,1983Pradhan-a,1983Pradhan-b}, it was shown by \citet{1982Pradhan,1985Pradhan-a} that, in particular, the triplet/singlet ratio G (see the definition below) depends considerably on the temperature and the ionization state of the plasma. Laboratory studies of ionization and recombination dominated plasmas \citep{1984Kallne,1985Lee,1986Lee} verified the expected behavior of G(T). This has also been discussed recently, among others, by \citet{2000Liedahl,2000Porquet,2000Bautista}.

\subsection{He-like iron}

 Earlier works on He-like iron have described different aspects of spectral variations. For example, \citet{1978Mewe-b} considered the time dependence of line ratios and ionization fractions, but not the detailed spectrum including DR satellites. In a later study \citep{1980Mewe} they examined time dependent spectra of a number of ionization states of Fe and Ca under solar flare conditions, without an external radiation field.  The effect of photoexcitation was first pointed out by \citet{1972Blumenthal}. \citet{1993Swartz} calculated the full \ka complex, but not the time dependent behavior of individual features. The present numerical simulations include all the features mentioned using new atomic data.

Recently, emission from the 6.7 keV complex of He-like Fe~XXV has been analyzed in ionization equilibrium including the important contribution by dielectronic satellites to the \ka complex at T $<$ 6.7 keV possibly prevalent in photoionized plasmas \citep{2001Oelgoetz}. They consider the set of forbidden (f), intercombination (i), and resonance (r) lines, with the proper spectroscopic designation labeled as w,x,y,z corresponding to the 4 transitions to the ground level $1s^2 \ (^1S_0) \longleftarrow 1s2p (^1P^o_1), 1s2p (^3P^o_2), 1s2p (^3P^o_1), 1s2s (^3S_1)$ respectively.  In recombination dominated, non-coronal plasmas \eion recombination preferentially enhances the triplet (x,y,z) lines, and hence the ratio G = (x+y+z)/w depends on the effective level-specific recombination rates, and therefore the H-like to He-like ratio ($X_{\makebox{Fe~XXVI}}$/$X_{\makebox{Fe~XXV}}$). Generally,  G $\approx 1  \rightarrow $  collisional (coronal) ionization equilibrium; G $> 1  \rightarrow $ recombination dominated (photoionization or hybrid); G $< 1  \rightarrow $ ionization dominated. 

But in addition to the principal lines the dielectronic satellites (DES) are observed to be quite prominent in the laboratory \citep[e.g.][]{1992Beiersdorfer}, solar flares and corona \citep[e.g.][]{1998Kato}.  Therefore the \ka complex of He-like Fe~XXV needs to be studied using a model that includes the \en = 2 bound levels, as well as the KL\en autoionizing levels of He-like ions. Using our earlier time-independent model in the HELINE code, Fig. 1 from \citet{2001Oelgoetz} shows the \ka complex of Fe~XXV computed in coronal equilibrium at several temperatures.  At temperatures well below that of maximum abundance of Fe~XXV, T $<$ T$_m \approx 10^{7.4}$ K \citep{1992Arnaud}, the DES dominate \ka emission.  On the other hand, for T $>$T$_m$ the DES intensities are small relative to that of principal lines w,x,y,z, with large variations in between. In lower temperature plasmas (e.g. \cite{1984Kallne}) and photoionized sources we expect the centroid of the \ka complex of Fe~XXV to lie shortward of the w-line and show `redward broadening' of the otherwise 6.7 keV feature. 

\citet{1993Swartz} \citep[see also][]{2001Oelgoetz} further show that in coronal equilibrium the ratio G(T) for Fe~XXV is approximately a constant, independent of T. However, it is clear from Fig. 1 that the $n$ = 2 KLL DES would be largely blended in with the principal lines since many of the strongest ones lie in between the z and the w lines.  Therefore, following several earlier works such as \citet{1993Swartz}, \cite{1982Bely-Dubau,1998Kato}, Oelgoetz and Pradhan assume that the DES(KLL) intensity is merged with the (x,y,z) lines (see Fig. \ref{ssfigure} ), but may be measured relative to the w-line blended with the higher-$n > 2$ satellites.  They they re-define the G ratio, including DES, as

\begin{equation}\label{GDequ}
    GD \equiv (x+y+z+KLL) \ / \ (w+ \sum_{n>2}KLn)
\end{equation}
 
\citet{2001Oelgoetz} established the temperature variations of the \ka complex in equilibrium. They also plot GD(T) from equation \ref{GDequ} as a function of T and the ionization fraction \xh.  It is ascertained that GD(T) falls sharply at T $>$T$_m$, owing to the decrease DES emission. Nonetheless, the effect of departure from ionization equilibrium was estimated in the time-independent model in an ad hoc manner.  In this study we introduce temporal-temperature dependence explicitly.

\subsection{Time-dependent plasmas}
It is generally assumed that ionization/recombination rates are much slower than radiative-collisional rates. Therefore the temporal-temperature (t,\te) variations would manifest themselves via the ionization fractions $X_i$ obtained from the coupled ionization balance equations

\begin{eqnarray}\label{ionbal}
\frac{dX_i}{dt} & = & N_e(-C_i(T_e,\phi) X_i - \alpha^{DR}_{i}X_i + \alpha^{RR}_{i}X_i + (\alpha^{R}_{i+1}(T) \nonumber \\
& & +\alpha^{DR}_{i+1})X_{i+1}+C_{i-1}(T,\phi)X_{i-1})
\end{eqnarray}
\begin{eqnarray}\label{ionrates}
C_i(T,\phi) & = & N_e^{-1}c\int{\frac{\phi(h\nu)}{h\nu}\sigma_i(h\nu) d\nu} \nonumber \\  & & +C_i^{DI}(T)+C_i^{EA}(T)
\end{eqnarray}
where $\alpha^{DR}_{i}$ and $\alpha^{RR}_{i}$ respectively are the dielectronic and radiative recombination rates out of ionization state i, C$_i$ are the rate coefficients for collisional ionization, with
C$^{DI}_i$ and C$^{EA}_i$ as the direct and excitation-autoionization
contributions respectively, $\phi(h\nu)$ the radiation energy density, and $\sigma_i$ the photoionization cross section of ionization state i.
Three body recombination is neglected, as it is unimportant for densities
presented in this study.  Note that the ionization fractions would vary with the transient behavior of any of the three parameters: \te, \ne, or $\phi(h\nu)$.
 The radiation field intensity may be expressed in
terms of the conventional ionization parameter U, the ratio of the
local photon to electron densities \citep{1999Krolik}

\begin{equation}
U=\frac{1}{N_e}\int_{1.0 Ryd}^{\infty} \frac{\phi(h\nu)}{h\nu} d\nu
\end{equation}

Once the ionization balance equations are solved, and the fractional populations of each ionization state found, the level populations of the He-like ion are then obtained by solving
\begin{eqnarray}\label{spectra}
\frac{dN_i}{dt} & = & \sum_{j>i}{N_j A_{ji}} - N_i\sum_{j<i}{A_{ij}}+\alpha_i^{H}(T)X_{H}N_e \nonumber \\
& & -\alpha^i_{Li}(T)N_iN_e + N_e\sum_{j\neq i}{N_jq_{ji}^e(T)} \nonumber \\ 
& & -N_eN_i\sum_{j\neq i}{q_{ij}^e} + N_e\sum_{j\neq i}{N_jq_{ji}^{p}(T)} \nonumber \\
& & -N_eN_i\sum_{j\neq i}{q_{ij}^{p}} + N_e\sum_{j\neq i}{N_jq_{ji}^{\alpha}(T)} \nonumber \\ 
& & -N_eN_i\sum_{j\neq i}{q_{ij}^{\alpha}} +N_e C_i^{Li}(T)X_{Li} \nonumber \\
& & -N_eC_{H}^i(T)N_i + \sum_{j\neq i} {N_j \phi(h\nu_{ji})B_{ji}} \nonumber \\
& & -N_i\sum_{j\neq i}{\phi(h\nu_{ij})B_{ij}}
\end{eqnarray}
where the $A$'s are the radiative decay rate coefficients, $B$'s absorption and stimulated emission rate coefficients, $\alpha_i$'s the level specific recombination rate coefficients, $C$'s the collisional ionization rate coefficients, $q_{ji}^e$, $q_{ji}^{\alpha}$, and $q_{ij}^{p}$ the rate coefficients for excitation/dexcitation by electrons, protons, and $\alpha$ particles respectively.    

The problem is then further simplified by using effective rates, including cascade corrections to reduce the model to a seven level problem.  Unified level specific \citep{2001Nahar-b} rates, are supplemented with distorted wave type radiative recombination and dielectronic recombination rates calculated using the Los Alamos edition of Cowan's Atomic structure code, CATS \citep[see:][]{1981Cowan,1988Abdallah} and the ionization code GIPPER \citep[see:][]{2000Archer} where the more detailed unified level specific rates are more difficult to obtain accurately.  Electron impact excitation rates into and out of high lying states (states for which close coupling calculations including resonances are unavailable) are also calculated in a similar manner, again via the distorted wave approximation, using the Los Alamos collisional code ACE \citep[see:][]{1988Clark}.   

In addition, the satellite line intensities due to dielectronic recombination (DR) and inner shell excitation (from the ground state) are calculated using the expressions of
\citet{1972Gabriel}.
\begin{equation}\label{DES}
I_s^{DR}=4\pi^{3/2}a_0^3X_{He}N_eT_e^{-3/2}e^{\frac{-E_s}{kT_e}}\frac{g_s
 \Gamma_r
\Gamma_a}{\Gamma_a+\sum{\Gamma_r}}
\end{equation}
                                                                                
\begin{equation}\label{ISES}
I_s^{ISCE}=X_{Li}N_e q(T_e) \frac{\Gamma_r}{\Gamma_a+\sum{\Gamma_r}}
\end{equation}
where $\Gamma_r$ is the radiative decay rate coefficient, $\Gamma_a$ the total autoionization rate coefficient, $E_s$ is the energy of the transition, $g_s$ the statistical weight of the upper state, and $q(T_e)$ is the electron impact excitation rate out of the ground state and into the autoionizing state that produces line s.

Inner shell photoexcitation is taken care of in a straight forward manner using the expression:
\begin{equation}\label{ISPE}
I_s^{ISPE}=X_{Li}\phi(h\nu_s) \frac{c^3 g_s}{8\pi\nu_s^2 g_{g.s.}}\left(\Gamma_{g.s.\leftarrow s}\right)\frac{\Gamma_r}{\Gamma_a+\sum{\Gamma_r}}
\end{equation}
using the same variable definitions above.

It should be noted that all coupling between the autoionizing states that give
rise to these satellite lines is neglected.

\section{COMPUTATIONS}

As discussed above, first we solve the coupled ionization balance equations (\ref{ionbal} and \ref{ionrates}).  Collisional ionization along with both dielectronic and radiative recombination rates are calculated on the fly using the expressions of \citet{1992Arnaud}.  Photoionization rates are pretabulated for a number of radiation fields by integrating over the resonance averaged cross sections of \citet{1998Bautista} and then read in.  If a steady state solution is desired, the matrix representation of the coupled ionization balance equations is factored via Gaussian elimination and then solved by the standard LINPACK routines.  Time dependent ionization fractions are found by employing the Fehlberg fourth-fifth order Runge-Kutta \citep{1969Fehlberg} routine by \citet{1977Shampine}.  The initial ionization fractions are either provided to HELINE in the input file, or more commonly the initial ionization fractions are those of a steady state plasma, and as such are calculated.  The resulting equations are then integrated forward in time, allowing the routine by Shampine and Watts to handle the choice of time step. 

Once the ionization fractions have been calculated for a given time or set of steady state conditions, spectra can be produced by solving the coupled set of equations that arise from equation \ref{spectra}.  We assume that the time scale for the ionization and recombination rates is much longer than the time scale for the radiative rates, as such equation \ref{spectra} is always solved in a time independent manner.  The collisional excitation rates of \citet{2001Whiteford} and \citet{1985Pradhan-b}, the level specific unified recombination rates of \citet{2001Nahar-b} and radiative decay rates of \citet{1999Nahar-a} are used along with the \citet{1967Lotz} formula for inner shell ionization.  Furthermore, ionization and recombination out of the n=2 levels of the He-like species is neglected, following the model of \citet{1978Mewe-a}.  Again Gaussian elimination is employed to first factor the matrix representation of the system of coupled equations, and then the problem is solved using the LINPACK routines.  The population of each of the four states that give rise to the four He-like lines of interest (w, x, y, z) are then multiplied by the corresponding radiative decay rates to produce line intensities.   Satellite line intensities are calculated using equations \ref{DES}, \ref{ISES}, \ref{ISPE}.  The autoionization and radiative decay rates are taken from \citet{1997Pradhan}; the inner shell excitation rates of \citet{1995Kato} are used as well.

Once line intensities have been calculated, each line is given a Gaussian line shape (with a width determined by the electron temperature).  Component spectra arising from radiative decay of the He-like spectroscopic states, both inner shell electron impact excitation and photoexcitation of the Li-like ground state, and dielectronic recombination into the KLL autoionizing states of the Li-like ion are written out separately, along with their sum, the total spectra.  The sequence then repeats, either with the next steady state ionization balance calculation, or by integrating the time dependent ionization balance equations forward to the next point of interest.

\section{RESULTS AND DISCUSSION}

 We compare and discuss steady state and transient plasmas with respect to (i) ionization balance, and (ii) the 6.7 keV \ka spectral features.  Their relevance to the standard He quartet line ratios R = z/(x+y) and G = (x+y+z)/w  are 
pointed out (It is usual practice to denote the forbidden (f or
z),
intercombination (i or x+y), and resonance (r or w) lines of He-like
ions as
a `triplet'; however, for highly charged ions such as Fe~XXV where the
fine structure separation between x and y is manifest and the excitation
atomic physics is distinct for each line, it is more
appropriate to refer to them as a `quartet', as herein), but more general diagnostics are developed including (a) the effect of radiation fields on ionization balance and photoexcitation of levels, (b) high-density effects, and (c) spectral time-variations in ionization, equilibrium, or recombination dominated plasmas.

\subsection{Steady State Plasmas}

 Under collisional or photoionization equilibrium the ionization balance may be fixed as function of temperature. It is then straightforward to obtain the spectral intensities using HELINE with $dX_i/dt = 0$ in Eq. (3).  
\subsubsection{Ionization Balance}

As described in the Computations section, the first step in any calculation is to solve the coupled set of equations arising from equations \ref{ionbal} and \ref{ionrates}.  Results of this step as a function of electron temperature for two different radiation fields, a blackbody and a power-law, along with a purely collisional case ($J=0$), are presented in Fig. \ref{ionbalfig}.  The radiation fields chosen are purely exemplary, after a number of trials, so as to highlight typical features.  Each radiation field is determined by either a Planck function or a 
power-law with a specified photon index.

The first thing one might note is that the radiation field allows a given ionization state to become abundant at lower electron temperatures relative to the pure collisional case.  Additionally, the range of electron temperatures over which a given ionization fraction can be found can be significantly broadened.  It should also be noted that for a faster decaying radiation field, such as a blackbody, the radiation field tends to effect the lower lying ionization states, but not the most highly charged ionization states that depend most significantly on the high energy tail.  This effect can be seen by comparing the ionization fraction curves for Fe XXIII - Fe XXVII in the upper (blackbody) and the lowest (pure collisional) panels of Fig. \ref{ionbalfig}, with the middle panel (power-law with a photon index  $\alpha$ = -2.3).  Lastly it should be noted that by working in terms of U (dividing the radiation field by $N_e$) all the ionization and recombination terms are first order with respect to electron density ($N_e$) and as such the ionization balance curves (as a function of electron temperature) in equilibrium only depend directly on the value of U and the shape of the radiation field, not on the over all intensity of the radiation field, or on the electron density.

The present approach, however, is explicitly predicated on the
assumption that in some transient plasmas ionization equilibrium or
energy balance is not satisfied globally or locally. For example,
fusion plasmas with sudden external energy input, or in flaring situations in
astrophysics, may not be constrained by parameters such
as direct coupling of temperature and radiation fields, cases normally
dealt with conventional photoionization equilibrium codes. This is one
of the crucial differences of the present method with existing codes;
the present code HELINE is therefore applicable to transient phenomena in 
both the laboratory and astrophysical situations.
The value of the ionization
parameter U is computed a posteori to illustrate correspondence with
a local radiation and electron density,
but is not confined to them. In fact, no single
temperature may be defined that corresponds to an exact particle
or radiation field distribution. HELINE attempts to simulate this
decoupling in a variety of ways, while also representing
photoionization and/or collisional equilibrium.

\subsubsection{Spectra}

The first patterns shown in the spectra of Fig. \ref{ssfigure} (panels in the first 4 rows) are the approximately linear scaling of intensity with electron density, and the importance of satellite lines at lower temperatures (LHS column).  The former is due to the fact that collisional excitation, collisional ionization, and recombination are the principle mechanisms of producing ions in the excited state, and all are first order with respect to electron density.  The intensities of Li-like satellites produced by inner-shell collisional excitation is greater at lower temperatures for two reasons.  First one can see in equation \ref{DES} that electron temperature appears both in the denominator, and in an exponential term.  As the electron temperature decreases, the dielectronic satellite lines grow in strength owing to the dominance of the exponential term in this temperature range.  Secondly, the Li-like ionization fraction appears in equation \ref{ISES} and Li-like iron ionizes quickly as one can see from Fig. \ref{ionbalfig}.

In order to simulate the effect of a relatively intense radiation field, such as may correspond to flaring conditions, we choose a fairly large value of U $\sim$ 200.  The results are shown in the bottom two panels of Fig. (\ref{ssfigure}).  The basic effect is analogous to that due to high densities (discussed in the next section), i.e. to redistribute level populations. However, the detailed effects of radiative and collisional transitions are different. Radiative transitions strongly favor dipole allowed excitations with large oscillator strengths. It was first noted by \citet{1972Blumenthal} that the $2^3S \rightarrow 2^3P$ transition is the most likely to be affected by photoexcitation.  It should be noted however that these dipole allowed absorption, and stimulated emission transitions must compete in overall rate with the radiative decay rates in order to be important, thus unlike the ionization balance problem, U is not the important parameter.  In the case of the transitions mentioned above, this implies strong radiation fields for all electron densities.

In order to keep the U values, and hence the ionization balance in a physically reasonable state, the increment in the intensity of the radiation field was done in conjunction with an increase in electron density.  Of the four transitions mentioned above, the transition from the $2^3S_1$ level to the $2^3P_{2}$ level is preferred as it has the greatest B-coefficient. The effects of photoexcitation along the $2^3S_1 \rightarrow 2^3P_0$ transition would not ordinarily be visible, as there is no spectral line produced by the decay of the $2^3P_0$ level.  The other transitions have similar B-coefficients to each other, and as such the transitions $2^3S_1 \rightarrow 2^3P_{1,2}$ are radiatively excited. As one would expect, this results in a diminishing of the z line, and an enhancement of the intensities of predominately the x and to a lesser degree the y line. The net effect is that photoexcitation decreases the usual electron-density sensitive line ratio R = z/(x+y), while G is essentially unchanged since it is the ratio of the sum of all triplets to the singlet w line.

Figure \ref{ssfigure} also shows the effects of electron density on the spectra.  The general trend is that as $N_e$ increases, population is moved from the metastable $2^3S_1$ level to the higher levels within the $n=2$ complex.  As the $2^3P_{1,2}$ levels are closer in energy than the $2^1P_1$, the first observed effect is the transfer of population into the $^3P$ levels and, as its decay rate is faster, a preferential enhancement of the y line over the x line. While at lower densities the x line increases faster than the y line owing to the higher collisional excitation rate of $2^1P_2$ than $2^1P_1$ from the $2^3S_1$, as the electron density increases the $n=2$ complex moves closer to statistical equilibrium.  Since the y line ($2^3P_1 \rightarrow 1^1S_0$) and the w line ($2^1P_1 \rightarrow 1^1S_0$) have significantly faster decay rates than the x ($2^3P_2 \rightarrow 1^1S_0$) line, the x line will weaken relative to the others as equilibrium inside the $n=2$ complex is approached.  These collisional processes at high densities eventually render the z line undetectable as can be seen in Fig. \ref{ssfigure}.  Again the net effect is a decrease in the value of the line ratio R while leaving G essentially unchanged; also the behavior is different from the photoexcitation case.  Should the electron density continue to be increased, the levels would finally come into thermodynamic equilibrium which would quench the x line as well. In order to calculate accurate spectra at these densities however, our model would need to modified to include three-body recombination, and other plasma effects such as line broadening, which would become important for higher densities ($N_e>10^{20}$).  Nevertheless, the main effect of z-line quenching at high densities is amply demonstrated in Fig. \ref{ssfigure} and is a potential plasma diagnostic of sources such as laser produced fusion plasmas.

At sufficiently high densities the DES will also be affected, as collisional rates among closed spaced autoionization levels approach typical autoionization rates of 10$^{13-14}$ sec$^{-1}$ at electron densities in excess of 10$^{20}$ cm$^{-3}$.  We do not consider such regimes herein, but further work is in progress.

\subsection{Transient Plasmas}

Characterizing the spectral `signatures' of transient or non-steady state plasmas is the primary goal of this study.  We assume sample asymmetric time dependent functions for electron temperature and radiation field intensity with a faster rise and a slower decay (such as under flaring conditions), shown in the upper two panels of Fig. \ref{tdepionbalfig}.  For illustrative purposes we assume a constant electron density ($N_e$) of 10$^{8}$cm$^{-3}$, which is appropriate for the chosen time scale of a few thousand seconds \citep[a time scale such as what was observed by][]{2003Dewangan}.  If the density is much lower, the plasma is unable to respond on the time scale of a few thousand seconds.  If the density is much higher the plasma will tend toward equilibrium much more quickly.  Not only does this make for a quasi steady state plasma (which was not our objective) but it also becomes incredibly time consuming to solve in a time dependent manner as the time steps needed to accurately solve the ionization balance equations get inordinately small.  Three time dependent models were used.  In the first just electron temperature was varied (as in the top panel of Fig. \ref{tdepionbalfig}); there was no radiation field.  This is the collisional case; it's ionization fractions can be found in the middle panel of Fig. \ref{tdepionbalfig} and it's spectra in the left column of Fig. \ref{tdepspectrafig}.  The radiation field intensity was varied in the second case (as in the second panel of Fig. \ref{tdepspectrafig}) and the electron temperature was held constant at $T_e=10^{6.0}$.  This is the photoionized case; it's ionization fractions are in the fourth panel of Fig. \ref{tdepionbalfig}, and it's spectra in the middle column of Fig. \ref{tdepspectrafig}.  Both electron temperature, and radiation field intensity were varied in the third case.  This hybrid case give rise the the ionization fractions found in the bottom panel of Fig. \ref{tdepionbalfig}, and it's spectra in the right most column of Fig. \ref{tdepspectrafig}.  The results and diagnostics derived from these three cases is discussed in the next three subsections.

\subsubsection{The Transient Collisional Plasma}

Ionization balance in the transient plasma is dominated by ionization and recombination at different times, as shown in the middle panel (2nd from Top)
of Fig. \ref{tdepionbalfig}, and the left column of Fig. \ref{tdepspectrafig}. Initially (about the first 1500 seconds), the inner shell collisional excitation (ISCE) lines are very prominent due to an overabundance of Li-like Fe XXIV and a higher than usual electron temperature - hence higher collisional excitation rates.  The z line is also enhanced due to inner-shell collisional ionization ($1s^22s + e^{-} \longrightarrow 1s2s (^3S_1) + 2e^{-}$) during the early part of this ionizing regime.  The effect on the line ratios is a larger value of G(GD) than one would otherwise expect for a given temperature, due to enhanced z and satellite lines.  In addition
the G(GD) line ratios would be large because the w line may not be excited
very early during the ionizing phase. The ratio R would be similarly
affected, although resolving the x and y lines might be difficult in this case.  It might be noted that this scenario refers to astrophysical or laboratory plasma sources where the collisional and radiative rates are averaged over distribution functions: Maxwellian for particles and Planckian for the radiation field. The resultant spectra may be different in controlled experimental setups where the incident electron beam can be mono-energetic and tuned to scan through the $n = 2$ levels. Selective excitation by such a beam \citep[as in Electron-Beam-Ion-Traps (EBIT), e.g.][]{1992Beiersdorfer,2003Takacs} may show a different 
signature during the ionization phase via tuned inner-shell ionization of the 
Li-like ions.  The EBIT beam widths may be represented by a Gaussian whose
widths are sufficiently mono-energetic to excite a given set of levels
primarily by electron impact excitation. Such simulations have been done
for line ratios measured in EBIT spectra of Ne-like iron Fe~XVII 
\citep{2002Chen,2003Chen}, using new Breit-Pauli R-matrix (BPRM)
collisional cross sections, and
rates computed with several electron distribution functions. Similar
computations are under way for line ratios for several He-like ions
using new collisional and radiative BPRM data. As such, the EBIT
experiments provide another means of line diagnostics in plasmas
different from astrophysical sources.

 In middle panels in the LHS column of Fig. \ref{tdepspectrafig}, the 
transitional period produces spectra that show a combination of 
recombination and ionization, and look like the steady state spectra of 
Fig. \ref{ssfigure}. Indeed, at $\sim$ 1560s, the spectra reflect quasi steady state conditions with similar line intensities as in Fig. \ref{ssfigure}, although some differences such as enhanced inner shell excitation lines due to an overabundance of the Li-like ionization state are apparent.

As recombination time scales in transient sources are likely to be
longer than ionizing time scales, the most common case occurs in the late phases (bottom panels of Fig. \ref{tdepspectrafig}) when one can see the switch to a regime where recombination begins to dominate the plasma conditions. Several features manifest themselves, driven mainly by recombination-cascades to the upper levels of the \en = 2 \ka complex. (I) The non-dipole x line becomes stronger than the y inter-combination line ($>$1600 seconds). (II) The forbidden z line is also enhanced in intensity. (III) As one can see from Fig. \ref{tdepionbalfig} the plasma becomes more and more ionized relative to the steady state as the electron temperature continues to cool faster than the plasma can recombine.  This lagging forces the plasma into a regime where recombination will dominate spectral formation. (IV) As recombination dominates over collisional ionization and excitation, the level populations in the $^3P_2$ and the $^3S_1$ levels are increased relative to others since recombination-cascades preferentially populate the triplet levels \citep{1982Pradhan,1985Pradhan-a}. This results in the enhancement of the z and x lines that can be clearly seen from about 1900 seconds on wards.  The effect is so pronounced that the forbidden lines z and x actually become stronger than the two dipole allowed lines w and y. We then have a `spectral inversion' compared to the steady state or ionizing conditions shown in the other panels of Fig. \ref{tdepspectrafig}.   This inversion would lead to an enhanced G(GD) ratio, and to an R 
value less than 1.

\subsubsection{The Transient Photoionized Plasma}

 The fourth panel of Fig. \ref{tdepionbalfig} and the middle column of Fig. \ref{tdepspectrafig} shows results for a sample, transient, photoionized plasma. There are similarities with the collisional model in that it has distinctive ionizing and recombining phases and that the transitional stage 
between them blends the features, but the most outstanding aspect of
photoionized plasmas is that throughout the process the DR and ISCE satellite lines are 
weak; the exponential terms in eq. \ref{DES} and \ref{ISES} dominate and are too small to allow the satellites to form. Thus only the narrow satellite lines from photoexcitation (q, r and t) along with a weak y line and a strong w line produce the spectra in the ionizing phase.  This is a huge difference from the afore described collisional case (in the panels of the left column).  This results in a blue-ward shift of the average flux and a G(GD) ratio which is much closer to 1 for much of the early ionizing phase.
 
The plasma then begins a gradual transition, as 
recombination becomes more important with increasing time.  At $\sim$1080 
seconds the plasma is just entering the recombination phase, as is evident 
by the now stronger non-dipole lines populated by recombination-cascades. 
The line intensities (in this case the He-like line intensities only) are
comparable and the spectra looks qualitatively like equilibrium spectra, but
without satellite lines.  At later times however, the plasma becomes 
dominated by recombination, and by about 1900 seconds the z and x lines have become stronger than the dipole allowed w and y lines. By about
2400 seconds the spectra of this transient photoionized case looks very 
similar in shape to the spectra of the transient collisional case.

\subsubsection{The Transient Hybrid Plasma}

When both the electron temperature and the radiation field intensity are 
varied in a time dependent manner, one would expect to see a mixture of the 
features of the previous two cases. The bottom panel of Fig. \ref{tdepionbalfig} and the right hand column of Fig. \ref{tdepspectrafig} shows such a situation.

First the plasma passes through an ionizing phase t $<$ 1500 seconds.
 During this time the w line is markedly enhanced along with the satellite lines produced by inner shell photoexcitation; the z line, 
along with the satellite lines arising from inner shell collisional excitation, are
stronger than in the photoionized case, but not dominate as in the purely collisional case. This indicates that both 
collisional and photoionization are ongoing, although photoionization is 
the more important of the two, as can be seen by the prominence of the w and q lines.  One need only look at the ionization fractions presented in the bottom panel of Fig. \ref{tdepionbalfig} to see that the radiation field presented is more effective at ionizing 
than the ramping electron temperature used.

By t $\sim$ 1560 seconds the plasma has made the transition into a recombining 
mode, although it has not yet begun to show the characteristic spectral 
inversion. One note of interest is that because the electron temperature 
is higher than in the transient photoionized case, the DES lines 
j and k can be seen, even if the principal He-like lines are more intense.  
By t $\sim$ 1920 seconds the spectral inversion that is common to all three 
models at late times is evident.

\section{CONCLUSION}

High resolution steady state and time dependent analysis of the 
\ka~complex of the primary (w,~x,~y, \& z) lines and dielectronic satellites
 is carried out in unprecedented detail. It spans 
an energy range between 6.6-6.7 keV; with a signature sensitive to ionization,
recombination, temperature, density, and radiation field. 

  The salient features of the simulated spectra are:

\begin{enumerate}
\item The dielectronic satellites are prominent around 6.65 keV at low 
temperatures, such as in 
some photoionized plasmas, shifting the \ka complex intensity red-ward to energies 
below the w line at 6.7 keV.
In general lower energy features may be present during both the
ionizing and recombining phases, although the latter is more likely.
\item Photoexcitation and electron density sensitive level redistribution 
affects the x and y lines differently;
photoexcitation tends toward a stronger x line, while electron density 
towards a stronger y line, and as such they may be able to be 
distinguished if the spectra are adequately resolved.\\
\item A time dependent source such as an X-ray flare can induce a number of 
observable spectral changes:
\begin{enumerate}
\item Enhancement of the z line and select satellite lines (q, r, t, u and v) due to inner shell collisional excitation lines during 
collisional ionization, resulting in higher than usual G(GD) ratio in
particular.
\item Enhancement of the resonance (w) and select satellite lines (primarily q, r, and t) during photoionization, 
resulting in a lower G(GD) ratio before the H-like Fe~XXVI ionization
fraction, and recombination-cascades therefrom, becomes important.
\item Enhancement of the non-dipole (z and x) lines and the `spectral
inversion' with respect to the dipole allowed y and w lines during
 recombination.
\end{enumerate}
\item In order to realize the full potential of the line diagnostics 
developed in this work, high resolution observations are needed.
This should be possible in X-ray spectroscopy from {\it Astro-E2} that is expected to resolve the Fe~XXV 6.7 keV complex with up to 6 eV resolution \citep{2003Yaqoob}. The basic diagnostics should also be applicable to inertial confinement fusion devices with nanosecond time scales, but considerably higher densities than in astrophysical sources.  Such simulations are in progress.\\
\end{enumerate}

\section*{Acknowledgments}
All of the modeling work and much of the close coupling data work was carried out at the Ohio Supercomputer Center in Columbus Ohio.  JO would also like to extend a special thanks to Honglin Zhang and Chris Fontes of Los Alamos National Laboratory (X-5) for their instruction in the use of the codes that produced the distorted wave data used in the radiative cascade part of the problem, and to Joe Abdallah, Jr. of Los Alamos National Laboratory (T-4) for the computational resources used to do those distorted wave calculations.  This work was partially supported by the NASA Astrophysical Theory Program (AKP) as well as partially conducted under the auspices of the United States Department of Energy (JO).

\onecolumn

\setlength{\unitlength}{1.0in}

\begin{figure}
\caption {Steady state ionization fractions for three different radiation 
fields as a function of electron temperature: top panel a blackbody ($B_{h\nu}(T_{h\nu} = 10^{6.5})$), middle panel --- power-law with $\alpha$ = -2.3, bottom panel --- collisional equilibrium without radiation field.  Both radiation fields correspond to same ionization parameter U $\sim$ 200. Notice that while the power law is not as effective at photoionizing lower ionization states, it has a greater effect on the higher ionization states due to it's slower decay.\label{ionbalfig}}
\begin{center}
\resizebox{\textwidth}{!}{
\begin{picture}(3.4,2.8)
\put(0.15,0.1){\resizebox{3.2in}{!}{\includegraphics{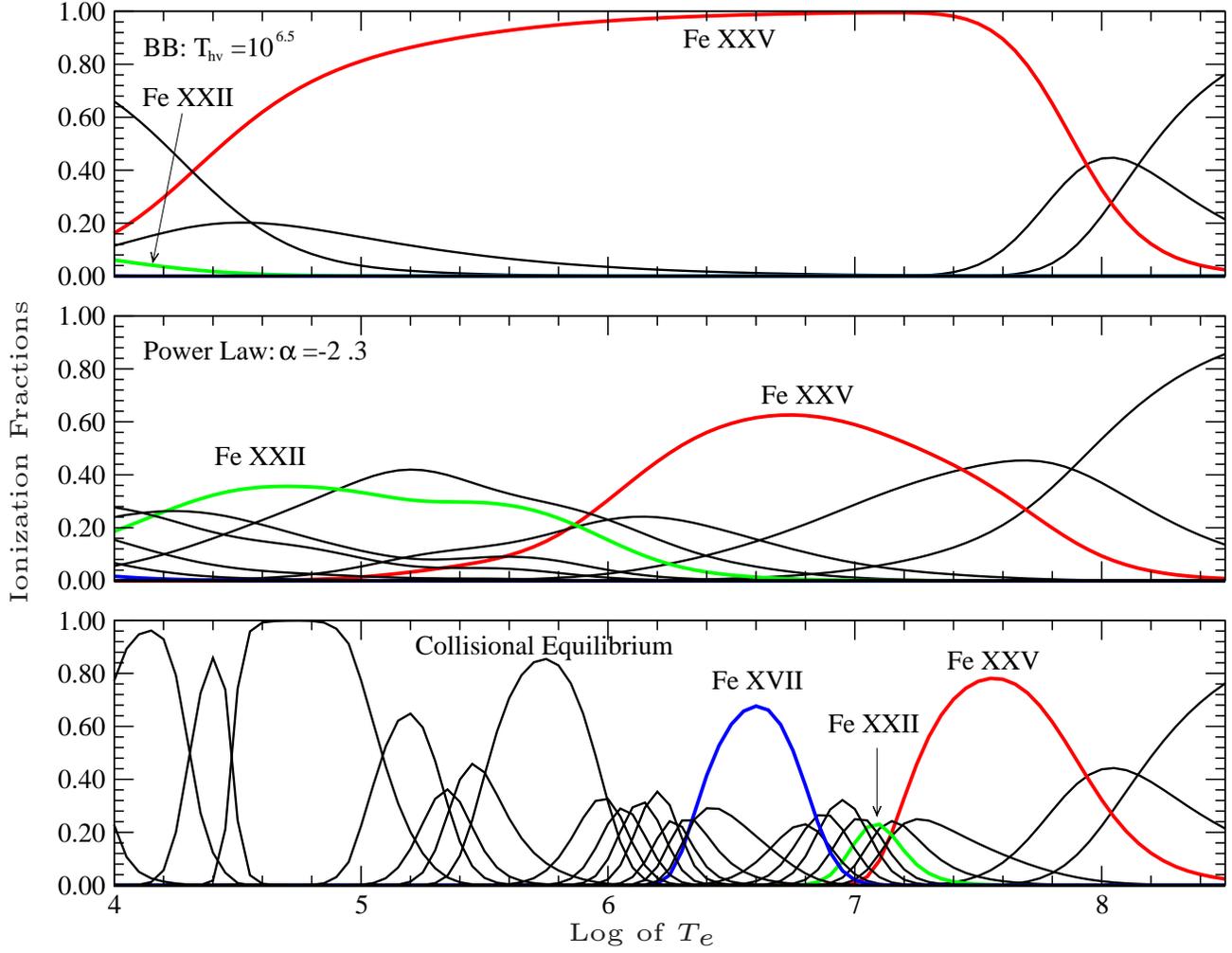}}}
\put(0.05,1.365){\rotatebox{90}{\makebox(0,0){\tiny Ionization Fractions}}}
\put(1.75,0.05){\makebox(0,0){\tiny Log of $T_e$}}
\end{picture}}
\end{center}
\end{figure}

\begin{figure}
\caption {Steady state spectra at a variety of densities and radiation fields. 
In particular, the spectra demonstrate the quenching of the forbidden
z line under different conditions.  All intensities are given in the same set of arbitrary, relative units.\label{ssfigure}}
\begin{center}
\resizebox{!}{0.9\textheight}{
\begin{picture}(7.1,8.8)
\put(1.1,0.6){\resizebox{6.0in}{!}{\includegraphics{fig2.eps}}}
\put(2.3,8.7){\makebox(0,0){\large $T_e=10^{7.4}$}}
\put(4.3,8.7){\makebox(0,0){\large $T_e=10^{7.7}$}}
\put(6.4,8.7){\makebox(0,0){\large $T_e=10^{8.0}$}}
\put(0.0,7.8){\makebox(0.9,0)[c]{\shortstack{\large Collisional\\\large$N_e=10^{10}$}}}
\put(0.0,6.6){\makebox(0.9,0)[c]{\shortstack{\large Collisional\\\large$N_e=10^{17}$}}}
\put(0.0,5.3){\makebox(0.9,0)[c]{\shortstack{\large Collisional\\\large$N_e=10^{18}$}}}
\put(0.0,3.9){\makebox(0.9,0)[c]{\shortstack{\large Collisional\\\large$N_e=10^{19}$}}}
\put(0.0,2.6){\makebox(0.9,0)[c]{\shortstack{\large $T_{h\nu}=10^{6.5}$\\\large$U \sim 200$\\\large$N_e=10^{17}$}}}
\put(0.0,1.3){\makebox(0.9,0)[c]{\shortstack{\large $T_{h\nu}=10^{6.5}$\\\large$U \sim 200$\\\large$N_e=10^{18}$}}}
\put(1.65,0.4){\makebox(5.45,0)[c]{\large Photon Energy in keV}}
\put(1.5,0.2){\resizebox{0.7in}{!}{\includegraphics{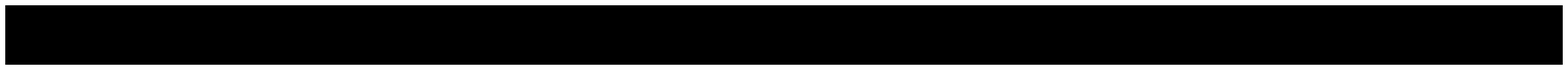}}}
\put(2.65,0.2){\resizebox{0.7in}{!}{\includegraphics{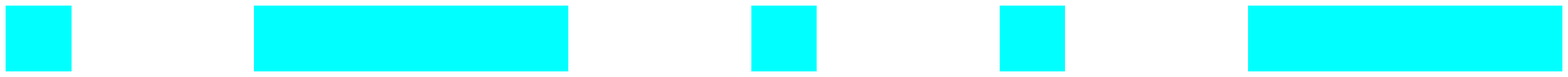}}}
\put(3.8,0.2){\resizebox{0.7in}{!}{\includegraphics{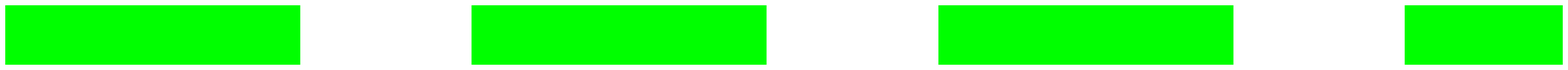}}}
\put(4.95,0.2){\resizebox{0.7in}{!}{\includegraphics{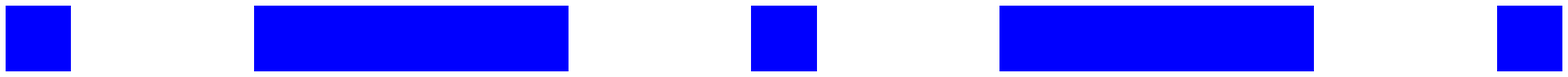}}}
\put(6.1,0.2){\resizebox{0.7in}{!}{\includegraphics{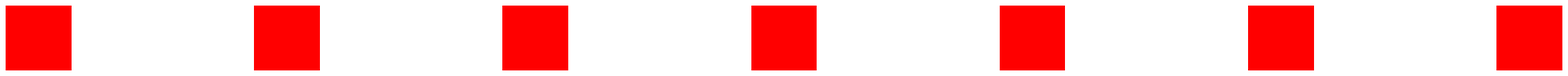}}}
\put(1.5,0.05){\makebox(0,0)[l]{Total Spectra}}
\put(2.65,0.05){\makebox(0,0)[l]{ISPE Satellite Lines}}
\put(3.8,0.05){\makebox(0,0)[l]{ISCE Satellite Lines}}
\put(4.95,0.05){\makebox(0,0)[l]{DR Satellite Lines}}
\put(6.1,0.05){\makebox(0,0)[l]{He-like Lines}}
\end{picture}}
\end{center}
\end{figure}

\begin{figure}
\caption{Time dependent profiles: Top: the electron temperature, 2nd from 
Top: ionization parameter U(t) associated with the radiation field, 
3rd from Top: ionization fractions under collisional ionization alone (T$_e(t)$
as above), 2nd from Bottom: 
ionization fractions for the transient photoionized case (variation of the radiation field intensity, U(t) as above and constant $T_e=10^{6.0}$), and Bottom: 
ionization fractions from the hybrid case (varying both T$_e(t)$ and U(t)) \label{tdepionbalfig}}
\begin{center}
\resizebox{!}{0.85\textheight}{
\begin{picture}(6.2,8.6)
\put(0.2,0.31){\resizebox{6.0in}{!}{\includegraphics{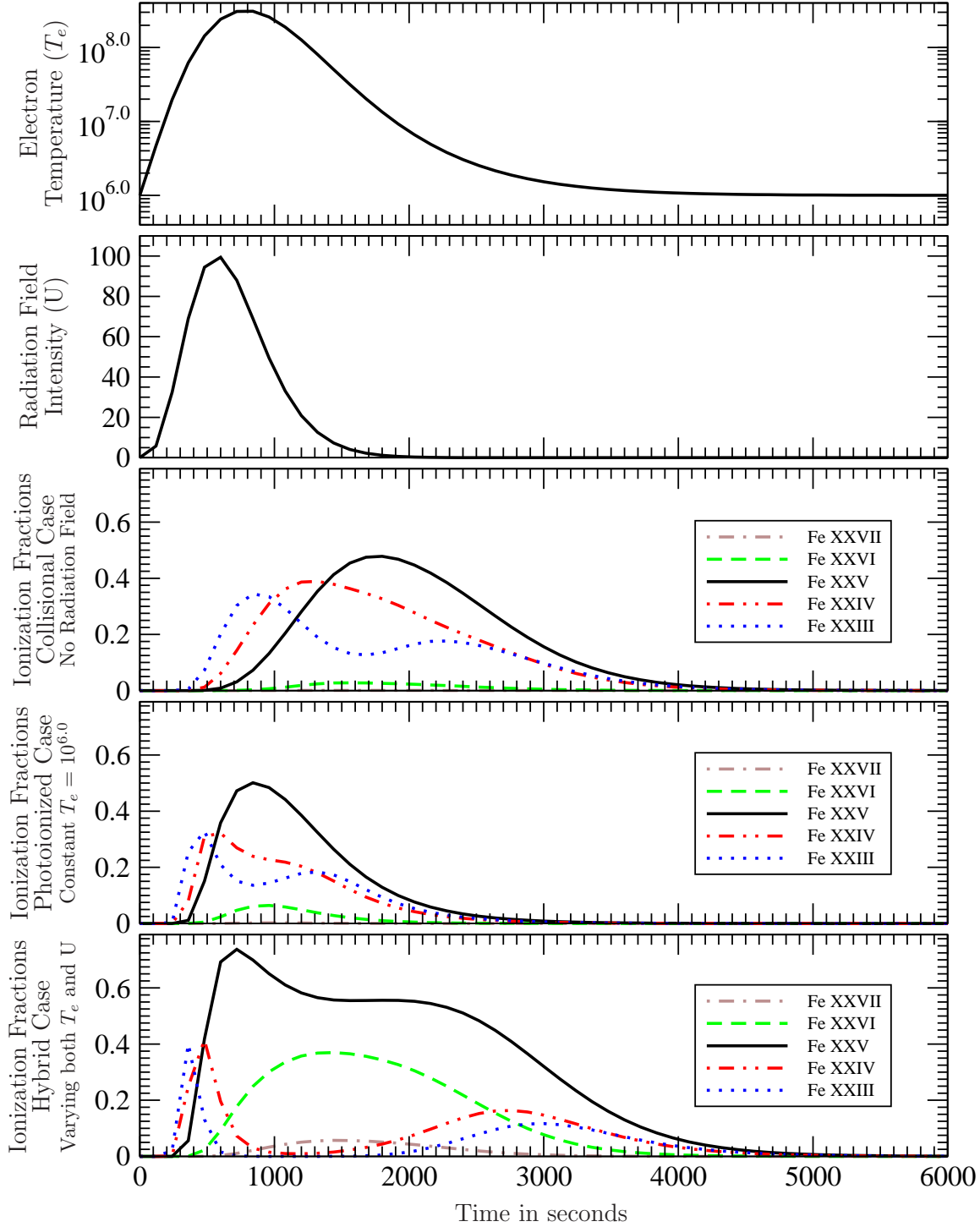}}}
\put(0.0,7.0){\rotatebox{90}{\makebox(0.9,0)[c]{\shortstack{\Large Electron\\\Large Temperature $(T_e)$}}}}
\put(0.0,5.45){\rotatebox{90}{\makebox(0.9,0)[c]{\shortstack{\Large Radiation Field\\\Large Intensity (U)}}}}
\put(0.0,3.9){\rotatebox{90}{\makebox(0.9,0)[c]{\shortstack{\Large Ionization Fractions \\ \Large Collisional Case \\\large No Radiation Field}}}}
\put(0.0,2.35){\rotatebox{90}{\makebox(0.9,0)[c]{\shortstack{\Large Ionization Fractions \\ \Large Photoionized Case \\ \large Constant $T_e=10^{6.0}$}}}}
\put(0.0,0.8){\rotatebox{90}{\makebox(0.9,0)[c]{\shortstack{\Large Ionization Fractions \\ \Large Hybrid Case \\ \large Varying both $T_e$ and U}}}}
\put(3.3,0.15){\makebox(0,0)[c]{\Large Time in seconds}}
\end{picture}}
\end{center}
\end{figure}

\begin{figure}
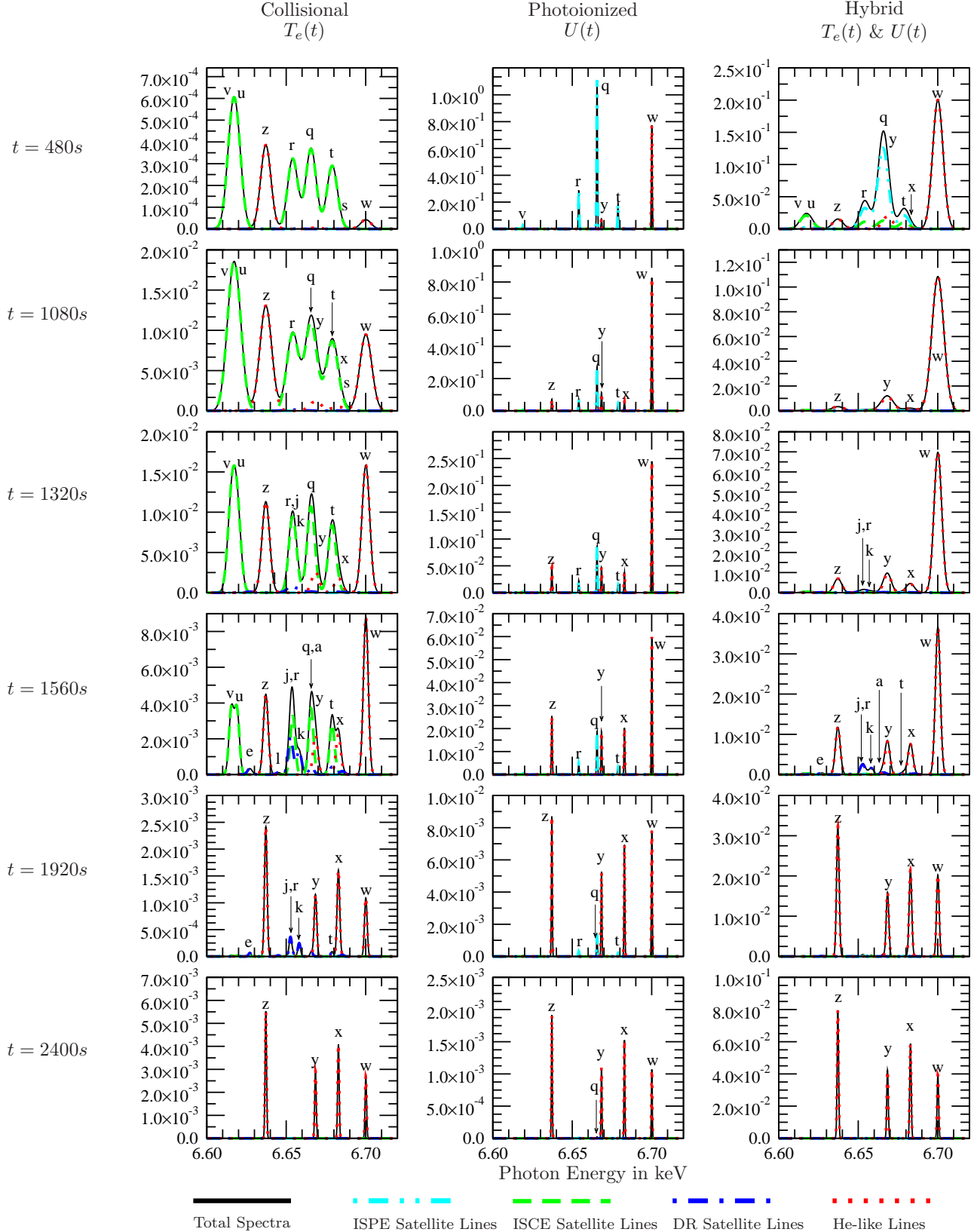

\caption{Evolution of transient spectra at various times throughout 
the time-dependent runs, with T$_e(t)$ and U(t) as in 
Fig. \ref{tdepionbalfig} (relative intensities).
The LHS panels correspond to pure collisional case, the middle panels to
photoionized case, and the RHS panels to the hybrid case.
 Of particular note are the enhanced satellite and z lines at early times
in the collisional case (signature of collisional ionization), 
the enhanced resonance (w) and inner shell photoexcitation (q,r \& t) lines at early times in the photoionized case and 
hybrid cases (signature of ongoing photoionization), and the enhanced 
non-dipole (x and z) lines at late times in all cases 
(signature of recombination).  All intensities are given in the same set of arbitrary, relative units.\label{tdepspectrafig}\label{lastpage}}
\begin{center}
\resizebox{!}{0.9\textheight}{
\begin{picture}(7.1,8.8)
\put(1.1,0.5){\resizebox{6.0in}{!}{\includegraphics{fig4.eps}}}
\put(2.3,8.7){\makebox(0,0){\shortstack{\large Collisional \\ \large $T_e(t)$}}}
\put(4.3,8.7){\makebox(0,0){\shortstack{\large Photoionized \\ \large $U(t)$}}}
\put(6.4,8.7){\makebox(0,0){\shortstack{\large Hybrid \\ \large $T_e(t)$ \& $U(t)$}}}
\put(0.0,7.8){\makebox(0.9,0)[c]{\large $t=480 s$}}
\put(0.0,6.6){\makebox(0.9,0)[c]{\large $t=1080 s$}}
\put(0.0,5.3){\makebox(0.9,0)[c]{\large$t=1320 s$}}
\put(0.0,3.9){\makebox(0.9,0)[c]{\large$t=1560 s$}}
\put(0.0,2.6){\makebox(0.9,0)[c]{\large$t=1920 s$}}
\put(0.0,1.3){\makebox(0.9,0)[c]{\large$t=2400 s$}}
\put(1.65,0.4){\makebox(5.45,0)[c]{\large Photon Energy in keV}}
\put(1.5,0.2){\resizebox{0.7in}{!}{\includegraphics{legend-black.eps}}}
\put(2.65,0.2){\resizebox{0.7in}{!}{\includegraphics{legend-cyan.eps}}}
\put(3.8,0.2){\resizebox{0.7in}{!}{\includegraphics{legend-green.eps}}}
\put(4.95,0.2){\resizebox{0.7in}{!}{\includegraphics{legend-blue.eps}}}
\put(6.1,0.2){\resizebox{0.7in}{!}{\includegraphics{legend-red.eps}}}
\put(1.5,0.05){\makebox(0,0)[l]{Total Spectra}}
\put(2.65,0.05){\makebox(0,0)[l]{ISPE Satellite Lines}}
\put(3.8,0.05){\makebox(0,0)[l]{ISCE Satellite Lines}}
\put(4.95,0.05){\makebox(0,0)[l]{DR Satellite Lines}}
\put(6.1,0.05){\makebox(0,0)[l]{He-like Lines}}
\end{picture}}
\end{center}
\end{figure}


\begin{thebibliography}{}


\bibitem[\protect\citeauthoryear{{Abdallah}, {Clark}, \& {Cowan}}{{Abdallah} et.~al.}{1988}] {1988Abdallah}{Abdallah} J., {Clark} {R.~E.~H.}, \& {Cowan} {R.~D.}, 1988, Los Alamos Manual {LA}-11436-{M}, vol I. {\em Theoretical Atomic Physics Code Development I {CATS}: {C}owan {A}tomic {S}tructure {C}ode}, Los Alamos National Laboratory

\bibitem[\protect\citeauthoryear{{Archer}, {Clark}, {Fontes}, \& {Zhang}}{{Archer} et.~al.}{2000}]{2000Archer}{Archer} {B.~J.}, {Clark} {R.~E.~H.}, {Fontes} {C.~J.}, \& {Zhang} {H.~L.}, 2000, Los Alamos Unlimited Release {LA}-{UR}-00-5693, {\em {GIPPER} user manual}, Los Alamos National Laboratory

\bibitem[\protect\citeauthoryear{{Arnaud} \& {Raymond}}{{Arnaud} \&
  {Raymond}}{1992}]{1992Arnaud}
{Arnaud} M.,  {Raymond} J.,  1992, {A}p{J}, 398, 394

\bibitem[\protect\citeauthoryear{Bailey, Chandler, Slutz, Bennett, Cooper,
  Lash, Lazier, Lemke, Nash, Nielsen, Moore, Ruiz, Schroen, Smelser, Torres \&
  Vesey}{Bailey et~al.}{2002}]{2002Bailey}
Bailey J.~E.,  Chandler G.~A.,  Slutz S.~A.,  Bennett G.~R.,  Cooper G.,  Lash
  J.~S.,  Lazier S.,  Lemke R.,  Nash T.~J.,  Nielsen D.~S.,  Moore T.~C.,
  Ruiz C.~L.,  Schroen D.~G.,  Smelser R.,  Torres J.,    Vesey R.~A.,  2002,
  {P}hys. {R}ev. {L}ett., 86, 1

\bibitem[\protect\citeauthoryear{{Bautista} \& {Kallman}}{{Bautista} \&
  {Kallman}}{2000}]{2000Bautista}
{Bautista} M.~A.,  {Kallman} T.~R.,  2000, {A}p{J}, 544, 581

\bibitem[\protect\citeauthoryear{{Bautista}, {Romano} \& {Pradhan}}{{Bautista}
  et~al.}{1998}]{1998Bautista}
{Bautista} M.~A.,  {Romano} P.,    {Pradhan} A.~K.,  1998, {A}p{J}{S}, 118, 259

\bibitem[\protect\citeauthoryear{{Beiersdorfer}, {Phillips}, {Wong}, {Marrs} \&
  {Vogel}}{{Beiersdorfer} et~al.}{1992}]{1992Beiersdorfer}
{Beiersdorfer} P.,  {Phillips} T.~W.,  {Wong} K.~L.,  {Marrs} R.~E.,    {Vogel}
  D.~A.,  1992, {P}hys. {R}ev. {A}, 46, 3812

\bibitem[\protect\citeauthoryear{{Bely-Dubau}, {Faucher}, {Dubau} \&
  {Gabriel}}{{Bely-Dubau} et~al.}{1982}]{1982Bely-Dubau}
{Bely-Dubau} F.,  {Faucher} P.,  {Dubau} J.,    {Gabriel} A.~H.,  1982,
  {M}{N}{R}{A}{S}, 198, 239

\bibitem[\protect\citeauthoryear{{Blumenthal}, {Drake} \&
  {Tucker}}{{Blumenthal} et~al.}{1972}]{1972Blumenthal}
{Blumenthal} G.~R.,  {Drake} G.~W.~F.,    {Tucker} W.~H.,  1972, {A}p{J}, 172,
  205

\bibitem[\protect\citeauthoryear{{Chen} \& {Pradhan}}{{Chen} \& {Pradhan}}{2002}]{2002Chen}{Chen} G.~X., {Pradhan} A.~K., 2002, {P}hys. {R}ev. {L}ett. 89, 013202-1

\bibitem[\protect\citeauthoryear{{Chen}, {Pradhan}, \& {Eissner}}{{Chen}, {Pradhan} \& {Eissner}}{2003}]{2003Chen}{Chen} G.~X., {Pradhan} A.~K., {Eissner} W., 2003, {J}. {P}hys {B}, 36, 453

\bibitem[\protect\citeauthoryear{{Clark}, {Abdallah}, {Csanak}, {Mann}, \& {Cowan}}{{Clark} et~al.}{1988}]{1988Clark}{Clark} {R.~E.~H.}, {Abdallah} J, {Csanak} G., {Mann} {J.~B.}, \& {Cowan} {R.~D.}, 1988, Los Alamos Manual {LA}-11436-{M}, vol. II, {\em Theoretical Atomic Physics Code Development II {ACE}: {A}nother {C}ollisonal {E}xcitation {C}ode}. Los Alamos National Laboratory 

\bibitem[\protect\citeauthoryear{{Collin}, {Coup{\' e}}, {Dumont}, {Petrucci}
  \& {R{\' o}{\. z}a{\' n}ska}}{{Collin} et~al.}{2003}]{2003Collin}
{Collin} S.,  {Coup{\' e}} S.,  {Dumont} A.-M.,  {Petrucci} P.-O.,    {R{\'
  o}{\. z}a{\' n}ska} A.,  2003, {A\&A}, 400, 437

\bibitem[\protect\citeauthoryear{{Cowan}}{{Cowan}}{1981}]{1981Cowan}{Cowan} R.~D., 1981, {\em The Theory of Atomic Structure and Spectra}. {U}niversity of {C}alifornia {P}ress

\bibitem[\protect\citeauthoryear{{Dewangan} \& {Griffiths}}{{Dewangan} \&
  {Griffiths}}{2003}]{2003Dewangan}
{Dewangan} G.~C.,  {Griffiths} R.~E.,  2003, preprint(astro-ph/0312342)

\bibitem[\protect\citeauthoryear{{Fehlberg}}{{Fehlberg}}{1969}]{1969Fehlberg}
{Fehlberg} E.,  1969, Technical Report NASA-TR-R-315, {\em Low-order classical
  Runge-Kutta formulas with stepsize control and their application to some heat
  transfer problems.}
NASA Marshall Space Flight Center

\bibitem[\protect\citeauthoryear{{Gabriel}}{{Gabriel}}{1972}]{1972Gabriel}
{Gabriel} A.~H.,  1972, {M}{N}{R}{A}{S}, 160, 99

\bibitem[\protect\citeauthoryear{{Gabriel} \& {Jordan}}{{Gabriel} \&
  {Jordan}}{1969}]{1969Gabriel}
{Gabriel} A.~H.,  {Jordan} C.,  1969, {M}{N}{R}{A}{S}, 145, 241

\bibitem[\protect\citeauthoryear{{Iwasawa}, {Lee}, {Young}, {Reynolds} \&
  {Fabian}}{{Iwasawa} et~al.}{2004}]{2004Iwasawa}
{Iwasawa} K.,  {Lee} J.~C.,  {Young} A.~J.,  {Reynolds} C.~S.,    {Fabian}
  A.~C.,  2004, {M}{N}{R}{A}{S}, 347, 411

\bibitem[\protect\citeauthoryear{{K\"{a}llne}, {K\"{a}llne}, {Dalgarno},
  {Marmar}, {Rice} \& {Pradhan}}{{K\"{a}llne} et~al.}{1984}]{1984Kallne}
{K\"{a}llne} E.,  {K\"{a}llne} J.,  {Dalgarno} A.,  {Marmar} E.~S.,  {Rice}
  J.~E.,    {Pradhan} A.~K.,  1984, {P}hys. {R}ev. {L}ett., 52, 2245

\bibitem[\protect\citeauthoryear{{Kato}, {Fujiwara} \& {Hanaoka}}{{Kato}
  et~al.}{1998}]{1998Kato}    
{Kato} T.,  {Fujiwara} T.,    {Hanaoka} Y.,  1998, {A}p{J}, 492, 822

\bibitem[\protect\citeauthoryear{{Kato}, {Safronova}, {Shlyaptseva}, {Cornille}
  \& {Dubau}}{{Kato} et~al.}{1995}]{1995Kato}
{Kato} T.,  {Safronova} U.,  {Shlyaptseva} A.,  {Cornille} M.,    {Dubau} J.,
  1995, Technical Report NIFS-DATA-24, {\em {C}omparison of the satellite lines of H-like and He-like Spectra.}  National Institute for Fusion Science, Nagoya 464-01, Japan

\bibitem[\protect\citeauthoryear{Krolik}{Krolik}{1999}]{1999Krolik}
Krolik J.~H.,  1999, {\em Active Galactic Nuclei}.  Princeton University Press

\bibitem[\protect\citeauthoryear{{Lee}, {Lieber}, {Chase} \& {Pradhan}}{{Lee}
  et~al.}{1985}]{1985Lee}
{Lee} P.,  {Lieber} A.~J.,  {Chase} R.~P.,    {Pradhan} A.~K.,  1985, {P}hys.
  {R}ev. {L}ett., 55, 386

\bibitem[\protect\citeauthoryear{{Lee}, {Lieber}, {Pradhan} \& {Xu}}{{Lee}
  et~al.}{1986}]{1986Lee}
{Lee} P.,  {Lieber} A.~J.,  {Pradhan} A.~K.,    {Xu} Y.,  1986, {P}hys. {R}ev.
  {A}, 34, 3210

\bibitem[\protect\citeauthoryear{{Liedahl}}{{Liedahl}}{2000}]{2000Liedahl}
{Liedahl} D.~A.,  2000, in {Bautista} M.~A.,  {Kallman} T.~R.,   {Pradhan}
  A.~K.,  eds, Atomic Data Needs for X-ray Astronomy: Proceedings The
  completeness criterion in atomic modeling.
NASA Publications, can be found at:
  http://heasarc.gsfc.nasa.gov/docs/heasarc/atomic/, pp 151--159

\bibitem[\protect\citeauthoryear{{Lin}, {Johnson} \& {Dalgarno}}{{Lin}
  et~al.}{1977}]{1977Lin-b}
{Lin} C.~D.,  {Johnson} W.~R.,    {Dalgarno} A.,  1977, {P}hys. {R}ev. {A}, 15,
  154

\bibitem[\protect\citeauthoryear{{Lotz}}{{Lotz}}{1967}]{1967Lotz}
{Lotz} W.,  1967, {A}p{J}{S}, 14, 207

\bibitem[\protect\citeauthoryear{{Mewe} \& {Schrijver}}{{Mewe} \&
  {Schrijver}}{1978a}]{1978Mewe-a}
{Mewe} R.,  {Schrijver} J.,  1978a, {A\&A}, 65, 99

\bibitem[\protect\citeauthoryear{{Mewe} \& {Schrijver}}{{Mewe} \&
  {Schrijver}}{1978b}]{1978Mewe-b}
{Mewe} R.,  {Schrijver} J.,  1978b, {A\&A}, 65, 115

\bibitem[\protect\citeauthoryear{{Mewe} \& {Schrijver}}{{Mewe} \&
  {Schrijver}}{1980}]{1980Mewe}
{Mewe} R.,  {Schrijver} J.,  1980, {A\&A}, 87, 261

\bibitem[\protect\citeauthoryear{{Nahar} \& {Pradhan}}{{Nahar} \&
  {Pradhan}}{1999}]{1999Nahar-a}  
{Nahar} S.~N.,  {Pradhan} A.~K.,  1999, {A\&AS}, 135, 347

\bibitem[\protect\citeauthoryear{{Nahar}, {Pradhan} \& {Zhang}}{{Nahar}
  et~al.}{2001}]{2001Nahar-b}
{Nahar} S.~N.,  {Pradhan} A.~K.,    {Zhang} H.~L.,  2001, {P}hys. {R}ev. {A},
  63, 060701

\bibitem[\protect\citeauthoryear{{Ness}, {Schmitt}, {Burwitz}, {Mewe},
  {Raassen}, {van der Meer}, {Predehl} \& {Brinkman}}{{Ness}
  et~al.}{2002}]{2002Ness}
{Ness} J.-U.,  {Schmitt} J.~H.~M.~M.,  {Burwitz} V.,  {Mewe} R.,  {Raassen}
  A.~J.~J.,  {van der Meer} R.~L.~J.,  {Predehl} P.,    {Brinkman} A.~C.,
  2002, {A\&A}, 394, 911

\bibitem[\protect\citeauthoryear{{Oelgoetz} \& {Pradhan}}{{Oelgoetz} \&
  {Pradhan}}{2001}]{2001Oelgoetz} 
{Oelgoetz} J.,  {Pradhan} A.~K.,  2001, {M}{N}{R}{A}{S}, 327, L42

\bibitem[\protect\citeauthoryear{{Porquet} \& {Dubau}}{{Porquet} \&
  {Dubau}}{2000}]{2000Porquet}
{Porquet} D.,  {Dubau} J.,  2000, {A\&AS}, 143, 495

\bibitem[\protect\citeauthoryear{Pounds, Reeves, King \& Page}{Pounds
  et~al.}{2003}]{2003Pounds}  
Pounds K.~A.,  Reeves J.~N.,  King A.~R.,    Page K.~L.,
 preprint(astro-ph/0310257)

\bibitem[\protect\citeauthoryear{{Pradhan}}{{Pradhan}}{1982}]{1982Pradhan}
{Pradhan} A.~K.,  1982, {A}p{J}, 263, 477

\bibitem[\protect\citeauthoryear{{Pradhan}}{{Pradhan}}{1983a}]{1983Pradhan-a}
{Pradhan} A.~K.,  1983a, {P}hys. {R}ev. {A}, 28, 2113

\bibitem[\protect\citeauthoryear{{Pradhan}}{{Pradhan}}{1983b}]{1983Pradhan-b}
{Pradhan} A.~K.,  1983b, {P}hys. {R}ev. {A}, 28, 2128

\bibitem[\protect\citeauthoryear{{Pradhan}}{{Pradhan}}{1985a}]{1985Pradhan-b}
{Pradhan} A.~K.,  1985a, {A}p{J}{S}, 59, 183

\bibitem[\protect\citeauthoryear{{Pradhan}}{{Pradhan}}{1985b}]{1985Pradhan-a}
{Pradhan} A.~K.,  1985b, {A}p{J}, 288, 824

\bibitem[\protect\citeauthoryear{{Pradhan}, {Norcross} \& {Hummer}}{{Pradhan}
  et~al.}{1981a}]{1981Pradhan-b}
{Pradhan} A.~K.,  {Norcross} D.~W.,    {Hummer} D.~G.,  1981a, {P}hys. {R}ev.
  {A}, 23, 619

\bibitem[\protect\citeauthoryear{{Pradhan}, {Norcross} \& {Hummer}}{{Pradhan}
  et~al.}{1981b}]{1981Pradhan-c}
{Pradhan} A.~K.,  {Norcross} D.~W.,    {Hummer} D.~G.,  1981b, {A}p{J}, 246,
  1031

\bibitem[\protect\citeauthoryear{{Pradhan} \& {Shull}}{{Pradhan} \&
  {Shull}}{1981}]{1981Pradhan-a}  
{Pradhan} A.~K.,  {Shull} J.~M.,  1981, {A}p{J}, 249, 821

\bibitem[\protect\citeauthoryear{{Pradhan} \& {Zhang}}{{Pradhan} \&
  {Zhang}}{1997}]{1997Pradhan}
{Pradhan} A.~K.,  {Zhang} H.~L.,  1997, {J}. {P}hys {B}, 30, L571

\bibitem[\protect\citeauthoryear{{Rees} \& {M{\' e}sz{\' a}ros}}{{Rees} \&
  {M{\' e}sz{\' a}ros}}{2000}]{2000Rees} 
{Rees} M.~J.,  {M{\' e}sz{\' a}ros} P.,  2000, {Ap{J}{L}}, 545, L73

\bibitem[\protect\citeauthoryear{{Shampine} \& {Watts}}{{Shampine} \&
  {Watts}}{1977}]{1977Shampine}
{Shampine} L.~F.,  {Watts} H.~A.,  1977, in {Forsythe} G.~E.,  {Malcolm} M.~A.,
    {Moler} C.~B.,  eds, Computer Methods for Mathematical Computations.
Prentice-Hall, Englewood Cliffs, N.J.

\bibitem[\protect\citeauthoryear{{Swartz} \& {Sulkanen}}{{Swartz} \&
  {Sulkanen}}{1993}]{1993Swartz}
{Swartz} D.~A.,  {Sulkanen} M.~E.,  1993, {A}p{J}, 417, 487

\bibitem[\protect\citeauthoryear{{Tak{\' a}cs}, {Silver}, {Laming}, {Gillaspy}, {Schnopper}, {Brickhouse}, {Barbera}, {Mantraga}, {Ratliff}, {Tawara}, {Mak{\' o}nyi}, {Madden}, {Landis}, {Beeman}, {Haller}}{{Tak{\' a}cs} et~al.}{2003}]{2003Takacs} {Tak{\' a}cs} E., {Silver} E., {Laming} J.~M., {Gillaspy} J.~D., {Schnopper} H., {Brickhouse} N., {Barbera} M., {Mantraga} M., {Ratliff} L.~P., {Tawara} H., {Mak{\' o}nyi} K., {Madden} N., {Landis} D., {Beeman} J., {Haller} E.~E.,  2003, Nuc. Inst. and Meth. in Phys. Res. B, 205, 144

\bibitem[\protect\citeauthoryear{{Whiteford}, {Badnell}, {Ballance}, {O'Mullane}, {Summers} \& {Thomas}}{{Whiteford} et~al.}{2001}]{2001Whiteford} {Whiteford} A.~D., {Badnell} N.~R., {Ballance} C.~P., {O'Mullane} M.~G., {Summers} H.~P. \& {Thomas} A.~L., 2001, J. Phys. B, 34, 3179

\bibitem[\protect\citeauthoryear{{Yaqoob}, {George}, {Kallman}, {Padmanabhan},
  {Weaver} \& {Turner}}{{Yaqoob} et~al.}{2003}]{2003Yaqoob}
{Yaqoob} T.,  {George} I.~M.,  {Kallman} T.~R.,  {Padmanabhan} U.,  {Weaver}
  K.~A.,    {Turner} T.~J.,  2003, {A}p{J}, 596, 85

\end{thebibliography}
\end{document}